\documentclass[a4paper,11pt]{article}
\pdfoutput=1 

\usepackage{jinstpub} 
\usepackage[T1]{fontenc} 
\usepackage{subcaption} 
\usepackage{comment}

\usepackage{lineno}
\title{\boldmath Conceptual design and progress of transmitting $\sim$ MV DC HV into 4 K LHe detectors}

\author[a]{Zhuo Liang, \note{First author.}}
\author[a]{Fengbo Gu,}
\author[a]{Jiangfeng Zhou,}
\author[a,b,1]{Junhui Liao,\note{Corresponding author.}}
\author[c,2]{Yuanning Gao,}
\author[a]{Zhaohua Peng,}
\author[a]{Jian Zheng,}
\author[a]{Guangpeng An,}
\author[a]{Meiyuenan Ma,}
\author[a]{Lifeng Zhang,}
\author[a]{Lei Zhang,}

\author[d]{Xiuliang Zhao,}

\author[e]{Junfeng Xia,}

\author[f]{Gang Liu,}
\author[f]{Shangmao Hu,}

\affiliation[a]{China Institute of Atomic Energy, \\ Sanqiang Rd. 1, Fangshan district, Beijing, China, 102413.}
\affiliation[b]{Department of Physics, Brown University, \\ Hope St. 182, Providence, Rhode Island, USA, 02912.}
\affiliation[c]{School of Physics, Peking University, \\ Chengfu Rd. 209, Haidian district, Beijing, China, 10084.}
\affiliation[d]{School of Nuclear Technology, University of South China, \\ Changsheng West Rd.28, Hengyang, Hunan, China, 421009.}
\affiliation[e]{Shanghai Electric Cable Research Institute Co.,Ltd., \\ Jungong Rd.1000, Yangpu District, Shanghai, China, 200093.}
\affiliation[f]{Electric Power Research Institute, China South Grid, \\ Guangzhou, China, 510663.}

\emailAdd{liangzhuo\_w@163.com}
\emailAdd{junhui\_liao@brown.edu}

\abstract{A dual-phase TPC (Time Projection Chamber) is more advanced in characterizing an event than a single-phase one because it can, in principle, reconstruct the 3D (X-Y-Z) image of the event, while a single-phase detector can only show a 2D (X-Y) picture. As a result, more enriched physics is expected for a dual-phase detector than a single-phase one. However, to build such a detector, DC HV (High Voltage) must be delivered into the chamber (to have a static electric field), which is a challenging task, especially for an LHe detector due to the extremely low temperature, $\sim$ 4 K, and the very high voltage, $\sim$ MV (Million Volts). This article introduces a convincing design for transmitting $\sim$ MV DC into a 4 K LHe detector. We also report the progress of manufacturing a 100 kV DC feedthrough capable of working at 4 K. Surprisingly, we realized that the technology we developed here might be a valuable reference to the scientists and engineers aiming to build residential bases on the Moon or Mars.}

\begin{document}
\maketitle
\flushbottom

\section{Searching for rare events with liquid noble gas TPCs} \label{sec1LNGTPCs}

Thanks to quite a few technical advantages~\cite{APPEC2022}, LAr (Liquid Argon) and LXe (Liquid Xenon) TPCs are widely implemented in rare events searching experiments such as DM direct detection~\cite{LZ2022, PandaX2021, XENONnT-NR-2023, DarkSide20k17, DEAP3600Website} and neutrino-less double beta decay~\cite{EXO2019}. Any event registered in a TPC can reconstruct its 3D image: the X and Y coordinates can be obtained by the photosensors array on the same plane, and the Z position can be obtained by its drifting time under an external electric field. The electric field can also determine the fraction of the ionized electron-ion pairs being separated. In general, the higher the field, the greater the number of ionized particles being separated, therefore, the larger the signal size. Additionally, a higher field would reduce the chance of events piling up, thanks to a shorter drifting time. However, the higher field brings up more challenges in handling the HV system, half of which are exposed at room temperature and another half in a cryogenic environment. A trade-off is often made to balance physics benefits and engineering efforts. None of the LAr and LXe TPCs currently running in underground labs has applied an HV greater than 50 kV, and the drift field is often $\sim < $ 200 V/cm. While for an LHe (Liquid Helium) TPC, a drift field of up to 10 kV/cm is required to have a $\sim$ 50\%~\cite{Phan20} of separation for electron-ion pairs and a reasonable speed of 2 m/s for electron bubbles~\cite{Donnelly98}. A 1.5 (330) kg LHe would need a 0.5 (1.5) m size\footnote{A cylindrical shape detector, the diameter equals to the height.} TPC to fill up and a 0.5 (1.5) MV (Million Volts) to apply to have the required field.

\section{Applying $\sim$ MV into an LHe TPC} \label{sec2MVToLHeTPC}

\subsection{The ALETHEIA project} \label{sec2Subsec1TheALETHEIAProject}

ALETHEIA stands for A Liquid hElium Time projection cHambEr In dArk matter. The ALETHEIA detector will implement the arguably most competitive technology in the DM direct detection community, TPC (Time Projection Chamber); and will be filled with the arguably cleanest bulk material, LHe~\cite{ALETHEIA-EPJP-2023}. However, since nobody has built such a detector before, many R\&D programs should be launched to address the availability of the technology. In this paper, we mainly focuses on the 500 kV High Voltage (HV) system.

\subsection{Progress on the HV system}\label{sec2Subsec2ProgressHVSystem}

As mentioned, a drifting field of 10 kV/cm is required for an LHe TPC to separate enough electrons from ionized electron-ion pairs (by the incident particles) and have a reasonable drift speed for the separated electrons. To safely deliver $\sim$ MV DC into an LHe TPC, we have launched extensive discussions with experts in accelerator physics, cryogenics, HV cables production, and ultra-high voltage power transmission ($\sim$ MV, AC, 1000 A current) since the project officially launched in 2020. 
We split the HV R\&D into three tasks: (a) HV power supplies, (b) Delivering HV into a TPC, and (c) TPC running safely with the 500+ kV applied. We limit our discussions to task (b)  in this paper.

As mentioned in references~\cite{Tommasini11, GencogluCebeci08, RahalHuraux79}, there are two categories of HV delivering failures. (a) A breakdown happened on the isolation layer between the anode and cathode. The breakdown is mainly due to the factors such as electric (field strength), thermal, mechanical, chemical, radiation, and contamination. (b) Another is a flashover, a breakdown between the two electrodes through the electrode's surface (not the insolation layer between the electrodes). The distance between an anode and cathode, surface smoothness, and pollutions are the most critical factors that respond to a flashover. So far, no reliable models to precisely predict a breakdown or a flashover due to the lack of complete understanding of the mechanism of the failures. However, many practical engineering experiences exist to prevent them from happening~\cite{Tommasini11}.

For an LHe TPC, the HV field is only to provide a static field, and the current is as small as $ \sim $ 100 $\mu$A (The same for the LZ HV power supply~\cite{LZTDR17}), which is different from ultra-high voltage power transmission where the current is $\sim$ 1000 A. In addition, as mentioned in reference~\cite{ALETHEIA-EPJP-2023}, there would be no impurities in an LHe detector's bulk material. As a result, we believe that the breakdown and  flashover could be mitigated at a certain level thanks to the small current and the extremely clean bulk material.

\begin{figure}[!t]	 
	\centering
    \includegraphics[width=1.0\textwidth]{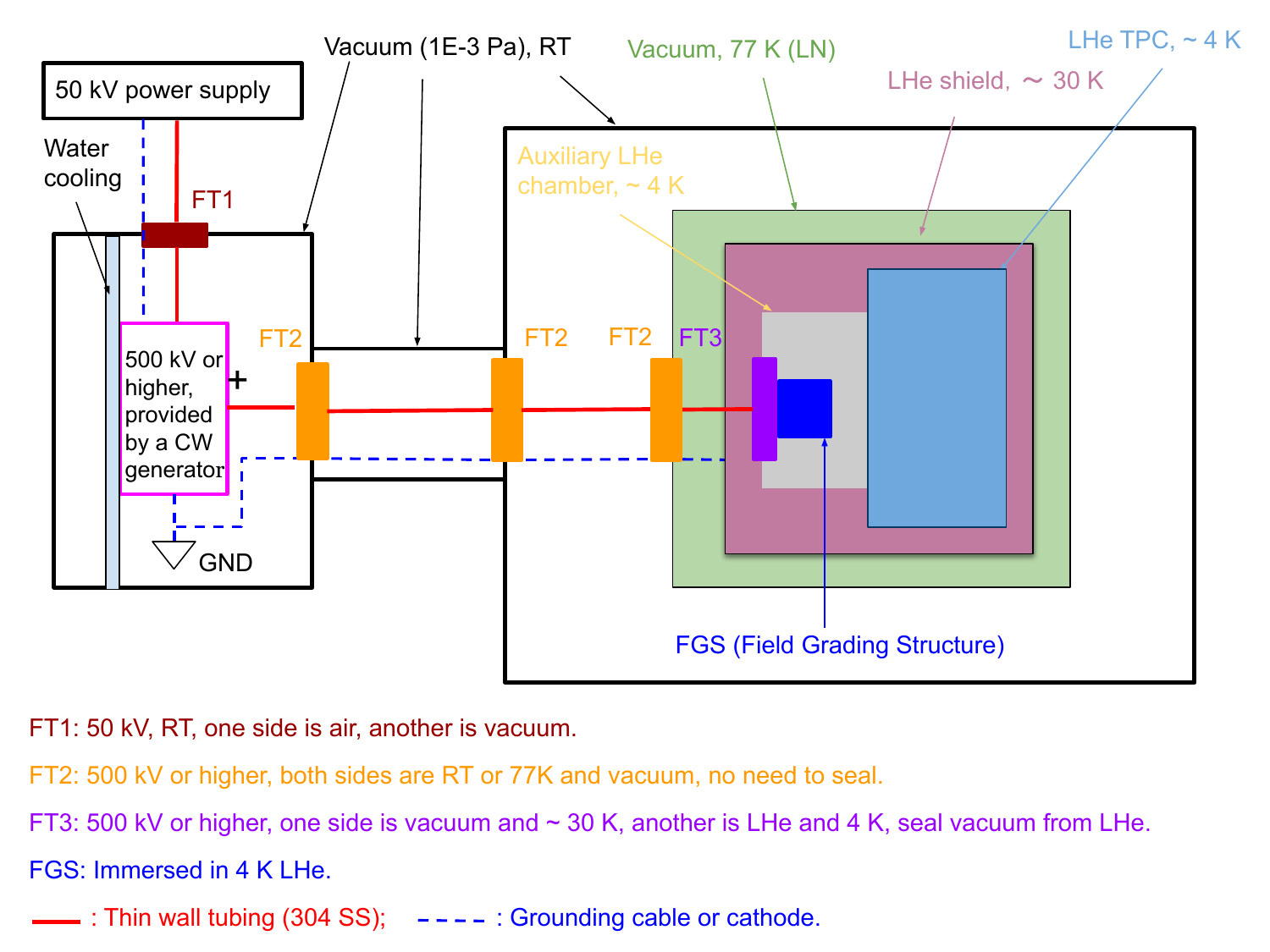}
	\caption{The preliminary version of our HV system. For details, please refer to the main text of the paper.}\label{scheme1SchematicDrawing} 
\end{figure}

Fig.~\ref{scheme1SchematicDrawing} shows schematically our preliminary design for HV transmission. A 50 kV power supply works at RT (Room Temperature) and 1 atm. Through the FT1, 50 kV can be transmitted into a RT and vacuum chamber, where 500 kV or higher HV can be generated via a commercial CW (Cockcroft-Walton) generator. Although the current of our HV is only $ \sim $ 100 $\mu$A, which means the heat generated by the HV system should be $\sim$ 100 Watts ($\approxeq$ 1 MV * 100 $\mu$A), a water cooling system is affiliated to get rid of the heat generated by the CW generator. The 500+ kV is transmitted down into the auxiliary LHe chamber via thin 304 SS tubings. Three FT2s anchor on the walls of variant chambers; since both sides of the FT2s are in a vacuum, they are not required to be sealed. Unlike FT2, which requires only 500+ kV isolation, FT3 has an additional requirement: sealing vacuum from LHe. FT3 is critical for the whole HV system and will be further discussed next. The FGS (Field Grading Structure) aims to degrade the local electric field on the liquid helium side of FT3. The HV transmission part of the scheme is similar to reference~\cite{Phan20}, which demonstrated that a 100 kV DC has run smoothly for 2 weeks at 0.5 K.

Almost all parts in our HV scheme are commercially available except FT2s and FT3. Moreover, since an FT3 can replace an FT2, we mainly focused on designing and making an FT3. Cross-linked polyethylene (XLPE) is widely implemented as isolation in commercial 500+ kV cables for ultra-high voltage power transmission~\cite{500kVXLPECable-IEC, 500kVXLPECable-China}, but it cannot be implemented into our FT3 since PE becomes crispy at LHe temperature according to our discussions with a few experts who experienced or knew for sure the PE crisp problem before. Polyimide (PI) possibly is the most suitable material since (a) it can work at a temperature down to 2 K ~\cite{DuPontVespelPartsShapesChinese, DuPontVespelDesignhandbook} and (b) has a (presumably) strong dielectric strength at 4 K. However, according to our searches, there is no data on a few cm thickness PI's dielectric strength at LHe temperature, though a 25 $\mu$m Kapton film's dielectric strength is measured to be 4100 kV/cm at LN (Liquid Nitrogen, 77 K) temperature~\cite{DuPontKaptonEng}. We also know that PI's dielectric strength decreases as long as thickness increases according to the tests at RT~\cite{DuPontVespelPartsShapesChinese}. Putting these pieces of information together, we estimate PI's dielectric strength at LHe temperature (4 K) is $\sim$ 100 kV/cm. We use the value as a reference to design a 100 kV FT and then test it to check whether the estimation is correct.

\subsection{The 100 kV DC feedthrough} \label{sec2Subsec3The100kVFT}

\begin{figure}[tbp]	 
	\centering
    \includegraphics[width=1.0\textwidth]{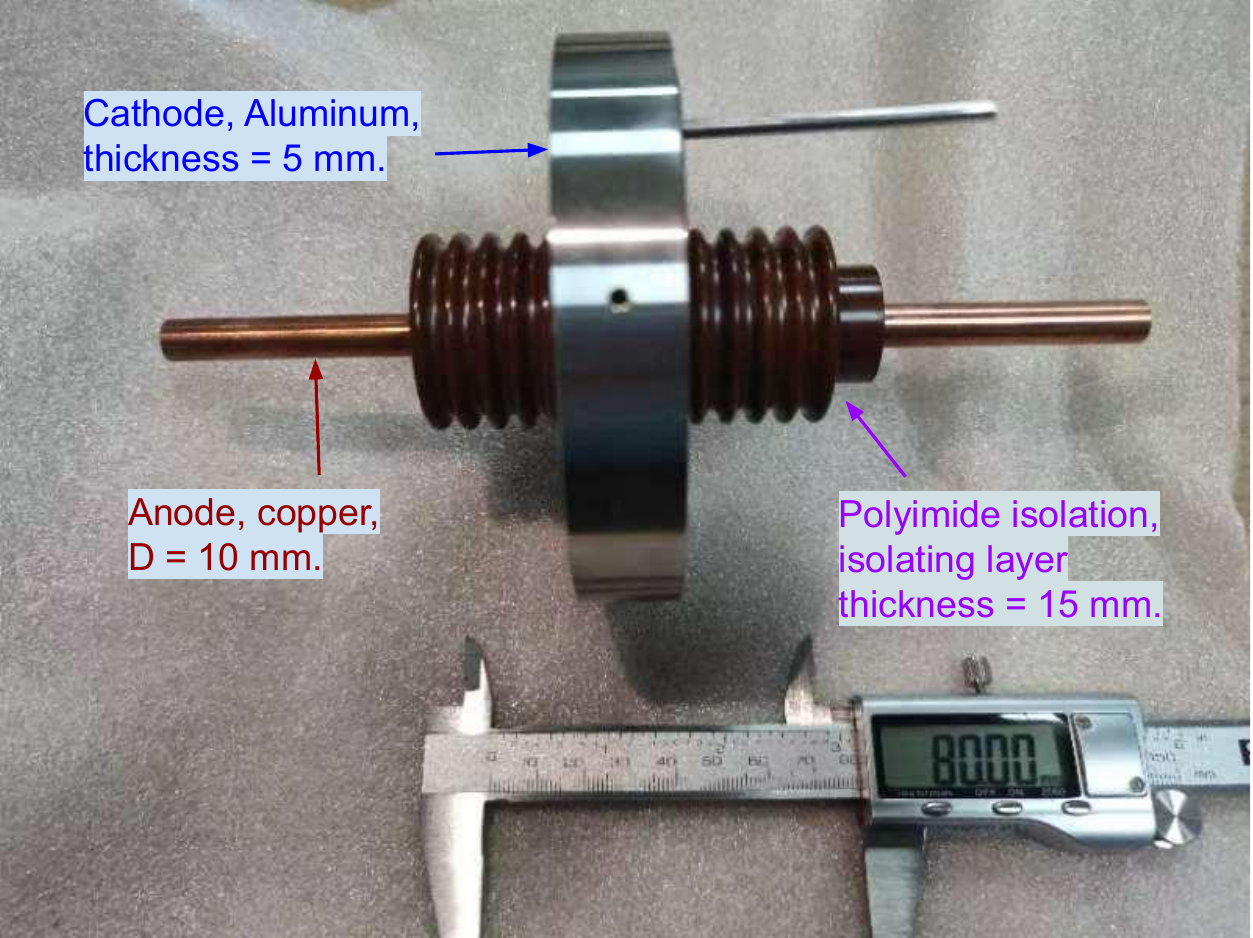}
	\caption{The picture of the FT we assembled at CIAE. The FT's anode is copper with the diameter of 10 mm. The isolating is made of polyimide, with a thickness of 15 mm. The cathode is 5 mm thick aluminum.}\label{fig100kVDCFT-1} 
\end{figure}

We designed an FT to convey 100 kV DC ($\sim$ 1 mA) at LHe temperature. Fig.~\ref{fig100kVDCFT-1} shows the picture of the house-made FT. The thickness of the polyimide layer is 1.5 cm, which should be enough to isolate 100 kV DC according to the assumed dielectric strength, 100 kV/cm. The surfaces of the isolating layer are polished to be roughness $< 0.8 ~\mu m$, and the surface distance between the anode and cathode is 8 cm. These efforts are supposed to decrease the possible flashover significantly. In addition, we carefully designed the machining process to eliminate mechanical stress during manufacturing and assembling.

\begin{figure}[!t]	 
	\centering
    \includegraphics[width=1.0\textwidth]{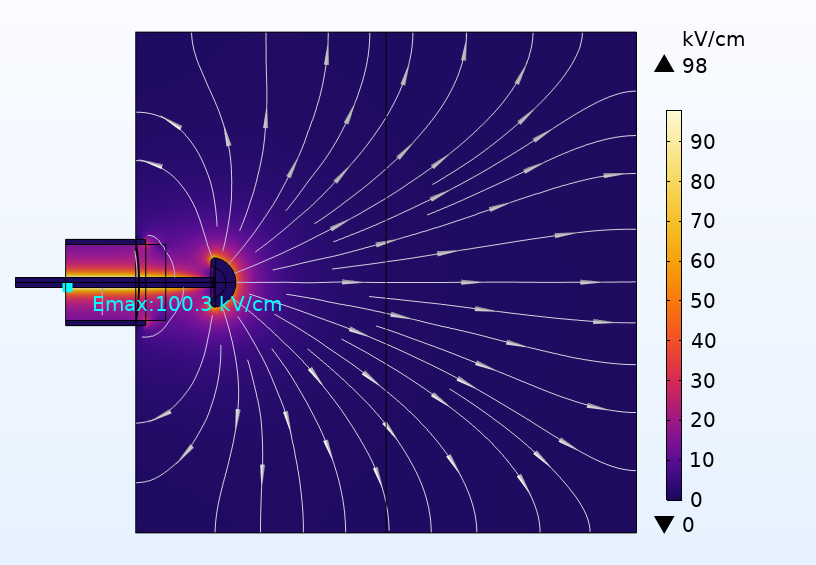}
	\caption{Simulating the electric field on the FT shown in~Fig~\ref{fig100kVDCFT-1} with COMSOL when 100 kV DC being applied to the anode. Inside of the dark purple area is LHe, outside is vacuum. The highest electric field is 100.3 kV/cm, roughly equals to the assumed breakdown voltage of polyimide at 4 K.}\label{fig100kVDCFT-2} 
\end{figure}

Simulations with COMSOL indicate the highest electric strength is 100 kV/cm, equivalent to the assumed polyimide's breakdown voltage at 4 K, as shown in Fig.~\ref{fig100kVDCFT-2}. Once the 100 kV FT passes the tests at a cryogenic temperature, we will make the FTs capable of working at 500 kV or higher voltages. The FT-related study will be summarized in a dedicated paper in the future.

Conventionally, there are roughly three isolating solutions for HV transmission. (a) In an HV coaxial cable, XLPE is used as an isolating material layer between the anode and cathode, (b) In a chamber, SF$_6$ is filled as an isolating medium, (c) for HV power long-distance transmission, air is implemented to isolate naked transmitting cables. In our scheme, we rely on polyimide and vacuum to isolate HV, as shown in Fig.~\ref{scheme1SchematicDrawing}. The HV goes through ambient air (RT) and a cryogenic vacuum in the scheme. 

It is well known that the surface of the Moon is a vacuum, and the highest and lowest surface temperatures are 140 $^\circ$C and - 248 $^\circ$C (25 K)~\cite{LunarTemp}, respectively. Like on Earth, the power should be transmitted from remote plants to residential bases, and there would be an interface (including feedthroughs) to convey HV from the vacuum, hugely varying temperature outer space into a room comfortable for human beings. Therefore, the interface is quite similar to our scheme because an HV has to go through the same two environments, ambient air (RT) and a cryogenic (and 140 $^\circ$C) vacuum.
As mentioned, the current ultra-high voltage transmission cable is based on XLPE~\cite{500kVXLPECable-IEC, 500kVXLPECable-China}; therefore, it can not work on the Moon. Giving that polyimide is functional from - 271 $^\circ$C (2 K) to 500 $^\circ$C ~\cite{DuPontVespelDesignhandbook}, it fits the environment on the Moon well.
However, we understand that to transfer electric power efficiently, electricity must be transmitted with a current of $\sim$1000 A; while our scheme aims at $\sim100 ~\mu$A, so our technology can not be transplanted directly for HV power transmission on the Moon; instead, additional R\&Ds are necessary to adapt the current. Nevertheless, our scheme can be a good starting point for the research.

\section{Summary}
HV might be the most challenging task to build a dual-phase LHe TPC since the detector requires an $\sim$ MV field for $\sim$ a meter size. After intensive discussions and conversations with experts in the related communities, we devised a convincing scheme to cope with the HV challenge: conveying 500+ kV DC from RT to 4 K temperature. We realized the technology might be a good reference to HV transmission in other extremely low-temperature circumstances, like building residential bases on the Moon or Mars.

\acknowledgments

Junhui Liao would also thank the support of the ``Yuanzhang'' funding of CIAE to launch the ALETHEIA program. This work has also been supported by NSFC (National Natural Science Foundation of China) under the contract of 12ED232612001001.



\bibliography{jinst2023}


\end{document}